\documentclass[12pt]{iopart}
\usepackage{iopams}  

\newcommand{\be}{\begin{equation}}
\newcommand{\ee}{\end{equation}}
\newcommand{\bd}{\begin{displaymath}}
\newcommand{\ed}{\end{displaymath}}
\newcommand{\ba}{\begin{array}}
\newcommand{\ea}{\end{array}}
\newcommand{\bt}{\begin{tabular}}
\newcommand{\et}{\end{tabular}}

\newcommand{\bea}{\begin{eqnarray}}
\newcommand{\eea}{\end{eqnarray}}
\newcommand{\bean}{\begin{eqnarray*}}
\newcommand{\eean}{\end{eqnarray*}}

\newcommand{\hlf}{\frac{1}{2}}

\newcommand{\cV}{{\cal V}}

\newcommand{\cM}{{\cal M}}

\newcommand{\dif}{\mathrm{d}}

\newcommand{\inp}[2]{\langle #1, #2 \rangle}

\newcommand{\R}{\mathbb{R}}

\newcommand{\ov}[1]{\overline{#1}}
\newcommand{\df}[1]{{} * \! #1}

\begin{document}
\begin{flushright}VUB-TENA-02-08\end{flushright} 

\title{Oxidation = group theory}

\author{Arjan Keurentjes\dag\ }

\address{\dag\ Theoretische Natuurkunde (TENA), Vrije Universiteit
  Brussel, Pleinlaan 2, B-1050 Brussels, Belgium}

\begin{abstract}
Dimensional reduction of theories involving (super-)gravity gives rise
to sigma models on coset spaces of the form $G/H$, with $G$ a
non-compact group, and $H$ its maximal compact subgroup. The reverse
process, called oxidation, is the reconstruction of the possible
higher dimensional theories, given the lower dimensional theory. In 3
dimensions, all degrees of freedom can be dualized to scalars. Given
the group $G$ for a 3 dimensional sigma model on the coset $G/H$, we
demonstrate an efficient method for recovering the higher dimensional
theories, essentially by decomposition into subgroups. The equations
of motion, Bianchi identities, Kaluza-Klein modifications and
Chern-Simons terms are easily extracted from the root lattice of the group
$G$. We briefly discuss some aspects of oxidation from the $E_{8(8)}/SO(16)$
coset, and demonstrate that our formalism reproduces the Chern-Simons
term of 11-d supergravity, knows about the T-duality of IIA and IIB
theory, and easily deals with self-dual tensors, like the 5-tensor of
IIB supergravity.
\end{abstract}

\ead{arjan@tena4.vub.ac.be}


\maketitle

\section{Introduction}

If one takes the idea of extra dimensions seriously, as is usually
done in Kaluza-Klein-, string- and M-theory, the immediate
question one faces is to guess the higher dimensional theory and
geometry from a lower dimensional one. The process of constructing
the higher dimensional theory from the lower dimensional one is called
``oxidation''. An interesting aspect of oxidation is that it is not
unique; there may be multiple theories giving rise to the same low
dimensional theory. This is the phenomenon of duality: Different
higher dimensional theories lead to the same low energy physics, and
we should not discriminate between the various higher dimensional
formulations. 

In this paper we discuss classical field theories only, and their
dimensional reduction (toroidal compactification and subsequent
truncation to the massless sector). For theories involving
(super-)gravity, this scheme leads to theories with scalar sectors
formulated as sigma models on symmetric spaces $G/H$, with $G$ a
non-compact group, and $G$ its maximal compact subgroup. Reducing to
2+1 dimensions, all bosonic degrees of freedom are scalars (or can be
dualized to scalars). We suppose that these live on a coset space
$G/H$, and ask the question: Which higher dimensional theories lead to
these theories, upon dimensional reduction ? 

Earlier studies of this question focussed on oxidation to 4 dimensions
  \cite{Breitenlohner:1987dg}, or to all dimensions, but restricted to
  cosets for which $G$ is split \cite{Cremmer:1999du}. One approach is
  to postulate the higher dimensional theory, and then derive that it
  indeed results in the wanted low dimensional theory. This is
  slightly unsatisfactory in view of the non-uniqueness of the
  oxidation process. A more systematic approach is presented in
  \cite{Henry-Labordere:2002dk}: Building on developments from
  \cite{Cremmer:1998px} and suggestive conjectures from
  \cite{Iqbal:2001ye} they developed a geometric scheme based on Del
  Pezzo-surfaces. A drawback is however that at present, this method
  only applies to groups that are split, and sub-groups of $E_{8(8)}$.

In \cite{Keurentjes:2002xc} we presented a scheme that allows one to
attack the problem for so-called split forms; the extension to generic
non-compact groups is discussed in \cite{Arjannew}. The method is
based on decomposition of $G$ into subgroups; due to its mathematical
nature it allows an exhaustive analysis. As a bonus we recover all
equations (equations of motion and Bianchi identities) for all
theories in the oxidation chain, and are able to furnish proofs of
some observations in the literature. As the method concentrates on
equations rather than actions, we have no difficulty in handling
theories with (anti-)self-dual tensors.

\section{The idea}

An important reference for the ideas expressed in this section is
\cite{Julia:1980gr}. For an extended discussion, and more references
we refer the reader to \cite{Keurentjes:2002xc}.

In theories of gravity the concept of a vielbein is an important one,
 and generalizations of the concept play an important role in coset
 theories. A vielbein represents an element of $GL(D)/SO(1,D-1)$. The
 quotient factor is the Lorentz group. It is standard lore that upon
 dimensional reduction, the number of (massless) degrees of freedom
 stays constant; these massless degrees organize in representations of the
 helicity group $SO(D-2)$. Motivated by this, we turn to the cosets
 $SL(D-2)/SO(D-2)$. We might consider $GL(D-2)$ instead of $SL(D-2)$,
 but the determinant of the $GL(D)$ element represents a scale factor that
 plays no role in the following. In 3 dimensions all degrees of
 freedom can be represented by scalars (after appropriate
 dualizations), and indeed, the dimensional reduction of general
 relativity in $D$ dimensions to $3$ dimensions gives rise to  a sigma
 model on $SL(D-2)/SO(D-2)$ \cite{Cremmer:1997ct}. 

We now consider 3 dimensional sigma models on coset spaces
of the form $G/H$, with $G$ a non-compact group, and $H$ its maximal
compact subgroup. A famous example is the 3-d coset on
$E_{8(8)}/SO(16)$, resulting from reducing the bosonic sector of a
maximal supergravity to 3 dimensions. This example immediately
suggests the inverse problem: Given the coset $G/H$ can this be
interpreted as a dimensional reduction of a higher dimensional theory
and if so, which one(s)? The result that general relativity gives rise
to a $SL(D-2)/SO(D-2)$ coset suggests that if the answer is
affirmative, we should be able to identify an $SL(D-2)$ subgroup in
$G$. Suppose we have such a subgroup, then we should decompose $G$ in
$SL(D-2)$ irreducible representations (irreps). Applying the
``vielbein'' we can convert $SL(D-2)$ irreps into $SO(D-2)$ irreps,
which in view of the previous discussion, we would like to interpret
as matter fields. 

The group $G$ is represented by its adjoint representation, which
is a self-conjugate irrep, and decomposition will lead to
self-conjugate representations. These come in two kinds: pairs of
conjugate irreps; and self-conjugate irreps. We want to interpret
these as massless fields, i.e. one graviton, forms and scalar
fields. The pairs of conjugate irreps represent forms and their duals;
self-conjugate irreps are the adjoint of $SL(D-2)$,
representing the graviton, singlets (scalars), or self-conjugate
forms. The latter are important in theories with (anti)self-dual tensors. 

The centralizer of $SL(D - 2)$ in $G$ acts as a symmetry group
 on the $SL(D-2)$ representations. We call the
 centralizer the \emph{U-duality group}, even though most cosets do
 not have a direct relation with string theory.

We have assumed thus far that upon decomposition of $G$ into $SL(D-2)$
irreps, we find irreps that can be interpreted as one (and no more)
graviton, and some forms and scalars. In other words, we want only one
adjoint irrep in the decomposition, and the rest must be
anti-symmetric tensors and singlets. Requiring this is equivalent to a
constraint on the $SL(D-2)$-subgroup in question; it has to be an
\emph{index 1} subgroup of $G$ (see \cite{Keurentjes:2002xc} for a
derivation of this fact). 

Index 1 subgroups are special. They must be \emph{regular}, which
means that that the root lattice of the subgroup can be chosen to be a
sublattice of the lattice of the original group. All regular subalgebra's
of a given (complexified) algebra can be found by a procedure
described by Dynkin \cite{Dynkin:um}. One still has to verify that the
algebra describes the real form that we want, $SL(D-2,\R)$. This can be
guaranteed by choosing a particular realization of the real form, and
some technology described in \cite{Keurentjes:2002xc, Arjannew}.   

If the group is simply laced, all regular embeddings have index 1. If
the group is non-simply laced, then regular subgroups involving only
short roots are possible. Such subgroups have an index bigger than
one, and hence are excluded.

The conclusion is that regular $SL(D-2, \R)$ subgroups of long roots are
the appropriate mathematical structure for encoding the graviton in
the theories under study. Regularity of subgroups is a criterion that
has been observed before \cite{Julia:1980gr}, and is implicit in many
discussions on algebraic aspects of compactified gravity. 
 
\section{Coset sigma models}

We use a basis for the Lie algebra of $G$ consisting of Cartan
subalgebra generators $H_i$ and ladder operators $E_{\alpha}$ (labelled
by the roots of $G$), obeying:
\be
\fl
[ H_i, H_j ]\!=\!0; \quad [ H_i, E_{\alpha}]\!=\!\alpha^i E_{\alpha}; \quad [
E_{\alpha}, E_{-\alpha}]\!=\!\alpha^i H_i; \quad [ E_{\alpha},
  E_{\beta}]\!=\!N_{\alpha,\beta} E_{\alpha+\beta}.
\ee 
Our discussion of sigma models is based on the action ($\cV \in G$)
\be
{L}_{G/H} =  -e \ \textrm{tr}\left((\partial \cV)
\cV^{-1}\hlf(1+T)(\partial \cV) \cV^{-1}\right) \label{taction}.
\ee
Here $T$ is an operator that encodes the action of the Cartan
involution on the algebra. As $T$ represents an involution,
$\hlf(1+T)$ is a projection operator, projecting out the generators of
the compact subgroup $H$. Although the analysis can be
done for any Cartan involution \cite{Arjannew}, we follow
\cite{Keurentjes:2002xc}, and restrict to $G$'s that are so called
split real forms. These are generated by linear combinations of the $H_i$'s
and $E_{\alpha}$'s with \emph{real} coefficients.

The 1-form $(\dif \cV)\cV^{-1}$ is an element of the algebra of
$G$. It can be expanded as
\be \label{tform}
(\dif \cV) \cV^{-1} = \hlf \sum_{i=1}^r \dif {\phi}^i {H}^i +
\sum_{\alpha \in  \Delta^{+}} e^{\hlf{\inp{\alpha}{\phi}}}
F_{(1)\alpha} E_{\alpha}.  
\ee
The second sum runs only over the set of positive roots of $G$,
$\Delta^+$, and $r$ denotes the (real) rank. This expansion assumes a
particular choice of gauge, the so-called positive root gauge, that is
possible by virtue of the Iwasawa decomposition. 

The $F_{(1)\alpha}$ are one forms. To get more detail on these, we
take the derivative $\dif((\dif \cV) \cV^{-1}) = ((\dif
\cV) \cV^{-1})\wedge((\dif \cV) \cV^{-1})$ which translates into
\be \label{scbi}
\dif F_{(1)\gamma} = \hlf \sum_{*} N_{\alpha,\beta} F_{(1)\alpha}
\wedge F_{(1)\beta} \qquad * = \left\{\ba{l} \alpha, \beta, \gamma \in
\Delta^+\\  
\alpha+ \beta = \gamma \ea \right..
\ee
Note the appearance of the structure constants $N_{\alpha,\beta}$. The
structure becomes even nicer when also considering the equations
of motion for the scalar coset. We recover these by regarding the
$F_{(1)\alpha}$ as independent fields, and enforcing
(\ref{scbi}) by Lagrange multipliers (which will be
($D-2$)-forms). Therefore we add to the Lagrangian 
\be
{L}_{Bianchi} = \sum_{\alpha \in \Delta^+}\left(\dif
F_{(1)\alpha} - \sum_{\alpha=\beta+ \gamma} N_{\beta, \gamma}
F_{(1)\beta} \wedge F_{(1)\gamma} \right)\wedge A_{(D-2)-\alpha} . 
\ee
The labels $-\alpha$ appearing on the $(D-2)$-forms turn out
meaningful. Varying with respect to $F_{(1)\gamma}$ we find the
equations  
\be \label{defmin}
F_{(D-1)-\gamma} \equiv e^{\inp{\gamma}{\phi}}\df{F}_{(1)\gamma} =
\dif A_{(D-2)-\gamma}-\sum_{\beta-\alpha=-\gamma} N_{\beta,-\alpha}
F_{(1)\beta}  \wedge A_{(D-2)-\alpha}. 
\ee
We defined $F_{(D-1)-\alpha}$, and used a group theoretical identity
for the structure constants. We are not interested in the
dual potential $A_{(D-2)-\alpha}$, but in the field strengths
$F_{(D-1)-\alpha}$. Taking the derivatives of these (note that in
general $\dif F_{(1)\beta} \neq 0$ !), one finds \cite{Keurentjes:2002xc} 
\be \label{sceqmo}
\dif F_{(D-1)-\gamma}= \sum_*  N_{\alpha,-\beta} F_{(1)\alpha} \wedge
F_{(D-1)-\beta}\qquad * =\left\{\ba{l} \alpha,\beta,\gamma \in
\Delta^+\\  \alpha-\beta=-\gamma \\ 
\ea \right. .
\ee
In terms of $F_{(D-1)-\alpha}$, the Lagrangian can be rewritten to
\be
{L}_{G/H} = - \inp{\df{\dif \phi}}{\dif \phi}-\hlf
\sum_{\alpha \in \Delta^+} F_{(D-1)-\alpha} \wedge F_{(1)\alpha},  
\ee
from which one finds the equation of motion for $\phi$ 
\be \label{cartaneqmo}
2 \dif(\df{\dif \phi}^i) = \sum_{\alpha \in \Delta^+} \alpha^i
F_{(D-1)-\alpha} \wedge F_{(1)\alpha}. 
\ee
Note that also the $\alpha^i$ are structure constants of the algebra.

With (\ref{scbi}),(\ref{sceqmo}) and (\ref{cartaneqmo}) we
have established a one-to-one relation between the conventional basis
of the Lie-algebra of $G$, and the equations relevant to the coset
sigma model. For every $E_{\alpha}$ where $\alpha$ is a positive root,
we have a Bianchi identity from (\ref{scbi}), when $\alpha$ is a
negative root we have an equation of motion from (\ref{sceqmo}), while
the Cartan subalgebra corresponds to equations (\ref{cartaneqmo}) for
$\phi$. The Cartan
subalgebra gives only $\textrm{rank }G$ equations, because the
``Bianchi identity'' for the potential for $\phi$, $\dif^2 \phi = 0$
is trivial. 

Coupling to other forms $F_{(n)}$ can be done by adding
quadratic terms to the action. If the $F_{(n)}$ form a non-trivial
representation of U-duality, we contract on an internal
``metric'' to render the action U-duality invariant. This schematically
takes the form  
\be \label{matteraction}
{L}_m = \frac{1}{2}\df{F}_{(n)} \wedge \cM F_{(n)},
\ee
with $\cM$ in an appropriate representation. The possibility of
Chern-Simons terms will turn up naturally later.   

The equation of motion, and Bianchi identity for $F_{(n)}$ can be
``covariantized'' to reflect the local $H$-invariance, and then become
\bea 
\left(\dif + T(\dif \cV \cV^{-1}) \right) \cV \df{F}_{(n)} = 0 & \quad &
\textrm{(equation of motion)}  \label{eqmo1}; \\
\left(\dif - (\dif \cV \cV^{-1})\right) \cV F_{(n)} = 0 & \quad &
\textrm{(Bianchi identity)}  \label{bi1}.
\eea
If the fields $\df{F}_{(n)}$ and $F_{(n)}$ represent the same
degrees of freedom, we cannot allow them to transform
differently. Hence, only local transformations $\cV(x) \rightarrow
U(x) \cV(x)$ for which $\hlf(1+T)(\dif U U^{-1})=0$ are allowed, but
this is precisely the restriction to the compact subgroup that was
imposed already.

Both $\cV F_{(n)}$ and $\cV \df{F}_{(n)}$ represent a full
$G$ multiplet of fields. The components of a multiplet can be
labelled by their weights, and we decompose by writing 
\be
\cV F_{(n)} \equiv \sum_{\lambda \in \Lambda}
e^{\hlf\inp{\lambda}{\phi}} F_{(n)\lambda}, \qquad  \cV \df{F}_{(n)}
\equiv \sum_{-\lambda \in \ov{\Lambda}} e^{-\hlf\inp{\lambda}{\phi}}
F_{(D-n)-\lambda}.   
\ee
The sum runs over the weights $\lambda$ ($-\lambda$) on the weight lattice
$\Lambda$ ($\ov{\Lambda}$) of the representation. $\cV \df{F}_{(n)}$
transforms in the representation conjugate that of $\cV F_{(n)}$, and
the weights differ by a minus sign.
The forms of degree $D-n$ are  \emph{not} the duals to
$F_{(n)\lambda}$. Rather,  
\be \label{id}
e^{-\hlf\inp{\lambda}{\phi}} \df{F}_{(D-n)-\lambda} =
e^{\hlf\inp{\lambda}{\phi}} F_{(n)\lambda}.  
\ee 

Inserting  $(\dif \cV) \cV^{-1}$ from (\ref{tform}) in the Bianchi
identity (\ref{bi1}), one obtains 
\be \label{bi2}
\dif F_{(n)\lambda'}= \sum_{*} N_{\alpha,\lambda} F_{(1)\alpha} \wedge
F_{(n)\lambda} \qquad * = \left\{\ba{l} \lambda, \lambda' \in
\Lambda\\ 
\alpha \in \Delta^+\\
\alpha+ \lambda = \lambda' \\
\ea \right. .
\ee
The constants $N_{\alpha,\lambda}$ can be computed but we will not need
them explicitly; when finding expressions like (\ref{bi2}) in the
future, the constants are determined in the derivation. 

The equation of motion (\ref{eqmo1}) can be rewritten similarly, and
becomes
\be \label{eqmo2}
\dif F_{(D-n)-\lambda'}= \sum_{*} N_{\alpha,-\lambda} F_{(1)\alpha}
\wedge F_{(D-n)-\lambda} \qquad * = \left\{\ba{l} \lambda,\lambda' \in
\Lambda \\ 
\alpha \in \Delta^+\\
\alpha-\lambda=- \lambda' \\
\ea \right. .
\ee

It can happen that form and dual form transform in a self-conjugate
representation; in theories with self-dual tensors they \emph{must} be
in such a representation. In that case the
equation of motion (\ref{eqmo2}) and Bianchi identity (\ref{bi2}) are
essentially the same equation, and we can consistently impose
self-duality. 

The Lagrangian (\ref{matteraction}) for coupled matter becomes
\be
{L}_m = - \frac{1}{2} \sum_{\lambda \in \Lambda}
F_{(D-n)-\lambda} \wedge F_{(n)\lambda} .
\ee
The sum over $\lambda$ indicates a sum over the weights of the
representation. Note that again there is one equation for every
weight.

With extra matter, the equation of motion for the dilatonic scalars
(\ref{cartaneqmo}) becomes 
\be \label{cartaneqmo2}
2 \dif(\df{\dif \phi}^i) = \sum_{\alpha \in \Delta^+} \alpha^i
F_{(D-1)-\alpha} \wedge F_{(1)\alpha} + \sum_{\lambda \in \Lambda}
\lambda^i F_{(D-n)-\lambda} \wedge F_{(n)\lambda}, 
\ee
while (\ref{sceqmo}) is not modified. Note that singlet
representations of U-duality do not couple to $\phi$ (as for these,
$\lambda=0$).   

For the non-dilatonic fields we have used roots and weights as
labels.  In equations we find that left and right hand side have the
same degrees, and their labels sum up to the same vector. The rule on
addition of forms is implied by Lorentz symmetry in the non-compact
directions. The additivity of the vector and weight labels follows
from the U-duality group. The theory is invariant under $\cV
\rightarrow U \cV$, with $U$ a constant element of $G$. Most of these
symmetries were eliminated by gauge fixing, but when $U$
is an element obtained by exponentiating an element of the Cartan
subalgebra, we have 
\be
\phi \rightarrow \phi + \zeta; \qquad F_{(n)\xi} \rightarrow
e^{-\hlf\inp{\xi}{\zeta}}F_{(n)\xi}.  
\ee
regardless of whether $\xi$ is a weight or a root. Covariance of the
equations of motion and Bianchi identities requires the
labels of various terms to add up on both sides. The exponential
factors included in the definitions of fields labelled by negative
roots (\ref{defmin}) and conjugate weights (\ref{id}) ensure the
transformation behavior implied by their labels.

\section{Oxidation}\label{inter}

The equations governing the axions (\ref{scbi}) (\ref{sceqmo}), and
form fields (\ref{bi2}) (\ref{eqmo2}) have a very similar common
structure. Together with the considerations on the group structure,
they suggest the following ``oxidation recipe'':
\begin{itemize}
\item We start from a 3 dimensional sigma model on $G/H$.
\item To oxidize to $D$ dimensions, decompose $G$ into $SL(D-2)
  \times U_D$ irreps, with $U_D$ the U-duality group, and $SL(D-2)$ an index
  1 subgroup. Of course, in the absence of such a group, we cannot
  oxidize. The irreps of $SL(D-2)$ are $n$-forms, singlets and one
  adjoint. The conjugate irrep of an $n$-form is a $(D-2-n)$-form.
\item Separate the roots of $G$ in $SL(D-2)$-weights (by projecting
  onto the subspace containing the $SL(D-2)$ lattice), and assign to
  each complete set of weights for a form representation a field
  $F_{(n+1)\alpha'}$, with $n$ the degree of the form, and $\alpha'$
  the projection of the root $\alpha$ onto the subspace orthogonal to
  the $SL(D-2)$ sublattice.
\item Some of the $SL(D-2)$ singlets are labelled by the roots of
  $U_D$ (these correspond to axions). Make a positive root decomposition
  for these, and assign a  form $F_{(1)\alpha}$ to positive roots, and
  $F_{(D-1)-\alpha}$ for negative roots.
\item \emph{All} (!) equations for the forms, and axions in the
  oxidized theories are given by 
\be \label{oxieq}
\dif F_{(n)\alpha'} = \hlf \sum_* \eta_{l,\beta;m,\gamma} \
N_{\beta,\gamma}  F_{(l)\beta'} \wedge F_{(m)\gamma'}  \quad * =
\left\{\ba{l} l+m=n+1\\  \alpha'+ \beta' = \gamma'
\ea \right. 
\ee
The $N_{\beta,\gamma}$ are structure constants inherited from $G$. We
  have included sign factors $\eta_{l,\beta;m,\gamma}= \pm 1$,
  because the structure constants are antisymmetric, but the product
  of two forms is not necessarily, and the sign factors prevent terms
  from vanishing pairwise. The computation of
  $\eta_{l,\beta;m,\gamma}$ is straightforward, and there is 
  freedom in the choice of signs \cite{Keurentjes:2002xc}. For many
  purposes it is not necessary to know them explicitly.
\item The remaining equations are the dilaton equation
  (\ref{cartaneqmo2}), and the Einstein equation. The terms on the
  right hand side of (\ref{cartaneqmo2}) and the matter fields to which
  the Einstein tensor couples are governed by the decomposition of $G$
  in $SL(D-2)$ irreps.
\end{itemize}
It can be demonstrated \cite{Keurentjes:2002xc} that this elegant and
simple procedure, is exactly the inverse to the systematic dimensional
reduction procedure as developed in \cite{Lu:1995yn, Cremmer:1997ct,
  Cremmer:1999du}.  Note that, whereas the equations
(\ref{scbi}),(\ref{sceqmo}),(\ref{bi2}) and (\ref{eqmo2}) always
involve a 1 form in the terms on the right side, we have made no such
restriction in (\ref{oxieq}). Such terms are known to arise due to modified
Bianchi identities, and Chern-Simons terms in the action.

\section{Examples}

Maximal supergravity in 3 dimensions gives rise to a sigma model on
$E_{8(8)}/SO(16)$. The group $E_{8(8)}$ has an index 1 subgroup $SL(9)$, under
which the 248-dimensional adjoint decomposes into the adjoint of
$SL(9)$, a 3-form, and a 6-form. The oxidation recipe gives rise to an
11 dimensional theory, with equations:
\bea
\dif F_{(7)} =  \hlf F_{(4)} \wedge F_{(4)} \qquad \dif F_{(4)} = 0
\eea 
The first equation is a consequence of the Chern-Simons term of
11-d supergravity.

There are two non-equivalent embeddings of $SL(8)$ into $E_{8(8)}$;
one results in an $SL(8) \times \R$ subgroup, and leads to the bosonic
sector of IIA supergravity; the other decomposition is $SL(8) \times
SL(2)$, giving the bosonic sector of IIB supergravity. All $SL(7)$
embeddings are equivalent, and can be embedded in the $SL(8)$ of IIA
theory or the one for IIB theory; this is the statement of T-duality
of IIA and IIB theory. 

Decomposing $E_{8(8)}$ into $SL(8) \times SL(2)$ we find the adjoints
of $SL(8)$ and $SL(2)$ (giving the graviton and a sigma model), an
$SL(2)$-doublet of $SL(8)$ 2-tensors, and an $SL(8)$ 4-tensor,
invariant under $SL(2)$. Equation (\ref{defmin}) and (\ref{oxieq})
result in
\be
\df{F}_{(5)0} = F_{(5)0} \qquad \dif F_{(5)0} = F_{(3)-\hlf\sqrt{2}}
\wedge F_{(3)\hlf\sqrt{2}} 
\ee
demonstrating that the formalism easily deals with self-dual tensors.

Our formalism covers a wealth of other theories, among which many that
allow a supersymmetric extension. More examples, and a discussion on
how the graphical language of Dynkin and Satake diagrams facilitates
the analysis can be found in \cite{Keurentjes:2002xc, Arjannew}. 
 
\ack I thank Niels Obers, Pierre Henry-Labordere
and Louis Paulot, and especially Bernard Julia for discussions. The
author is supported in part by the ``FWO-Vlaanderen'' through project
G.0034.02, in part by the Federal office for Scientific, Technical and
Cultural Affairs trough the Interuniversity Attraction Pole P5/27 and
in part by the European Commision RTN programme HPRN-CT-2000-00131, in
which the author is associated to the University of Leuven.

\end{document}